\documentclass[12]{article}
\usepackage{fleqn}
\usepackage{graphicx}
\begin{document}
\Large
\begin{center}
{\bf Sets of Mutually Unbiased Bases as Arcs in Finite Projective
Planes?}
\end{center}
\vspace*{-.3cm} \large
\begin{center}
Metod Saniga$^{\dag \ddag}$ and Michel Planat$^{\ddag}$
\end{center}
\vspace*{-.5cm} \small
\begin{center}
$^{\dag}${\it Astronomical Institute, Slovak Academy of Sciences,
05960 Tatransk\' a Lomnica, Slovak Republic}

\vspace*{.0cm}
 and

\vspace*{.0cm} $^{\ddag}${\it Institut FEMTO-ST, CNRS, Laboratoire
de Physique et M\'etrologie des Oscillateurs,\\ 32 Avenue de
l'Observatoire, F-25044 Besan\c con, France}
\end{center}

\vspace*{-.3cm} \noindent \hrulefill

\vspace*{.0cm} \small \noindent {\bf Abstract}

\noindent This note is a short conceptual elaboration of the conjecture of
Saniga {\it et al}  (J. Opt. B: Quantum Semiclass. {\bf 6} (2004)
L19-L20) by regarding a set of mutually unbiased bases (MUBs) in a
$d$-dimensional Hilbert space as an
analogue of an arc in a (finite) projective plane of order
$d$. Complete sets of MUBs thus correspond to ($d$+1)-arcs, i.e.,
ovals. In the Desarguesian case, the existence of two 
principally distinct kinds of ovals
for $d=2^n$ and $n \geq 3$, viz. conics and non-conics,
implies the existence of two qualitatively different groups of the
complete sets of MUBs for the Hilbert spaces of corresponding
dimensions. A principally new class of complete sets of MUBs are those having their analogues in
ovals in non-Desarguesian projective planes; the lowest dimension when this happens is $d=9$.

\vspace*{.15cm} \noindent {\bf Keywords:} 
Mutually Unbiased Bases, Ovals in (non-)Desarguesian Planes, Quantum Information Theory

\vspace*{-.1cm} \noindent \hrulefill

\vspace*{.3cm} \normalsize \noindent It has for a long time been
suspected but only recently fully recognized [1--4] that finite
(projective and related) geometries may provide us with important clues for
solving the problem of the maximum cardinality of MUBs for Hilbert 
spaces of finite dimensions $d$. It is well-known [5,6] that this number 
cannot be greater
than $d$+1 and that this limit is reached if $d$ is a power of a
prime. Yet, a still unanswered question is if there are
non-prime-power values of $d$ for which this bound is attained. On
the other hand, the minimum number of MUBs was found to
be three for all dimensions $d \geq 2$ [7]. Motivated by
these facts, Saniga {\it et al} [1] have conjectured that the
question of the existence of the maximum, or complete, sets of
MUBs in a $d$-dimensional Hilbert space if $d$ differs from a
prime power is intricately connected with the problem
of whether there exist projective planes whose order $d$ is not a
power of a prime. This note aims at getting a deeper
insight into this conjecture by introducing particular objects in
a finite projective plane, the so-called ovals, which can be
viewed as geometrical analogues of complete sets of MUBs.

We shall start with a more general geometrical object of a
projective plane, viz. a $k$-arc -- a set of $k$ points, no three
of which are collinear [see, e.g., 8,9]. From the definition it
immediately follows that $k=3$ is the minimum cardinality of such
an object. If one requires, in addition, that there is at least
one tangent (a line meeting it in a single point only) at each of
its points, then the maximum cardinality of a $k$-arc is found to
be $d$+1, where $d$ is the order of the projective plane [8,9];
these  ($d$+1)-arcs are called $ovals$. It is striking to observe
that such $k$-arcs in a projective plane of order $d$ and MUBs of
a $d$-dimensional Hilbert space have the {\it same} cardinality
bounds. Can, then, individual MUBs (of a $d$-dimensional Hilbert
space) be simply viewed as points of some abstract projective
plane (of order $d$) so that their basic combinatorial properties
are qualitatively encoded in the geometry of $k$-arcs? A closer
inspection of the algebraic geometrical properties of ovals
suggests that this may indeed be the case.

To this end in view, we shall first show that every proper
(non-composite) conic in $PG(2,d)$, a (Desarguesian) projective plane over the
Galois field $GF(d)$, is an oval. A conic is the curve of second
order
\begin{equation}
{\cal Q}:~ \sum_{i \leq j} c_{ij} z_i z_j = 0,~~i,j=1,2,3,
\end{equation}
where $c_{ij}$ are regarded as fixed quantities and $z_i$ as
variables, the so-called homogeneous coordinates of the projective
plane. The conic is degenerate (composite) if there exists a
change of the coordinate system reducing Eq.\,(1) into a form of
fewer variables; otherwise, the conic is proper (non-degenerate).
It is well-known [see, e.g., 8] that the equation of any {\it
proper} conic in $PG(2,d)$ can be brought into the canonical form
\begin{equation}
{\cal \widetilde{Q}}:~ z_1 z_2 - z_3^2 = 0.
\end{equation}
From the last equation it follows that the points of ${\cal
\widetilde{Q}}$ can be parametrized as $\varrho z_i=
(\sigma^2,1,\sigma), \varrho \neq 0$, and this implies that a
proper conic in $PG(2,d)$ contains $d$+1 points; the point
$(1,0,0)$ and $d$ other points specified by the sequences
$(\sigma^2,1,\sigma)$ as the parameter $\sigma$ runs through the
$d$ elements of $GF(d=p^n)$, $p$ being a prime and $n$ a positive
integer. Moreover, it can easily be verified that any triple of
distinct points of ${\cal \widetilde{Q}}$ are linearly independent
(i.e., not on the same line), as [10]
\begin{equation}
\det \left( \begin{array}{ccc}
1 & 0 & 0 \\
\sigma_1^2 & 1 & \sigma_1 \\
\sigma_2^2 & 1 & \sigma_2
\end{array}\right)
= \sigma_2 - \sigma_1 \neq 0
\end{equation}
and
\begin{equation}
\det \left( \begin{array}{ccc}
\sigma_1^2 & 1 & \sigma_1  \\
\sigma_2^2 & 1 & \sigma_2 \\
\sigma_3^2 & 1 & \sigma_3
\end{array}\right)
= (\sigma_1 - \sigma_2)(\sigma_2 - \sigma_3)(\sigma_3 - \sigma_1)
\neq 0.
\end{equation}
Hence, a proper conic of $PG(2,d)$ is indeed an oval. The converse
statement is, however, true for $d$ odd only; for $d$ even and
greater than four there also exist ovals which are {\it not}
conics [8--11]. In order to see this explicitly, it suffices to
recall that all the tangents to a proper conic ${\cal Q}$ of
$PG(2,d=2^n)$ are concurrent, i.e., pass via one and the same
point, called the nucleus [8--11]. So, the conic ${\cal Q}$
together with its nucleus form a ($d$+2)-arc. Deleting from this
($d$+2)-arc a point belonging to ${\cal Q}$ leaves us with an oval
which shares $d=2^n$ points with ${\cal Q}$. Taking into account
that a proper conic is uniquely specified by {\it five} of its
points, it then follows that such an oval cannot be a conic if $n
\geq 3$; for, indeed, if it were then it would have with ${\cal
Q}$ more than five points in common and would thus coincide with
it, a contradiction.

Let us rephrase these findings in terms of the above-introduced
MUBs~--~$k$-arcs analogy. We see that whilst for any $d=p^n$ there
exist complete sets (c-sets for short) of MUBs having their
counterparts in proper conics, $d=2^n$ with $n \geq 3$ also
feature c-sets whose analogues are ovals which are not conics. In
order words, our analogy implies that MUBs do not behave the same
way in odd and even (power-of-prime) dimensions. And this is,
indeed, the property that at the {\it number theoretical} level
has been known since the seminal work of Wootters and Fields [5,
see also 7], being there intimately linked with the fact that
so-called Weil sums
\begin{equation}
\left|\sum_{k\in GF(p^n)}e^{\frac{2\pi i}{p} {\rm Tr}(mk^2 + nk)}
\right|,
\end{equation}
with $m,n \in GF(p^n)$ and the absolute trace operator ``Tr"
defined as
\begin{equation}
{\rm Tr} (\eta) \equiv \eta + \eta^p + \eta^{p^{2}} + \ldots +
\eta^{p^{n-1}}, ~~ \eta \in GF(p^n),
\end{equation}
are non-zero (and equal to $\sqrt{p^n}$) for all $p>2$, playing
thus a key role for proving the mutual unbiasedness in these
cases, but vanish for $p=2$ [see, e.g., 12]. In the light of our analogy, this
difference acquires a qualitatively new, and more refined,
algebraic-geometrical contents/footing. Remarkably, this
refinement concerns especially even ($2^n$) dimensions, as we
shall demonstrate next.

In the example above, we constructed a particular kind of an oval
by adjoining to a proper conic its nucleus and then removing a
point of the conic; such an oval, called a pointed-conic, was
shown to be inequivalent to a conic for $n\geq3$. However, for
$n\geq 4$ there exists still another type of non-conic ovals,
termed irregular ones, that cannot be constructed this way [see,
e.g., 8,11,13]. This intriguing hierarchy of oval's types is
succinctly summarized in the following table:

\vspace*{0.2cm}
\begin{center}
\begin{tabular}{|c|cccc|}
\hline
$n$ &1&2&3& $\geq 4$ \\
\hline
ordinary conic & yes & yes& yes & yes \\
pointed-conic & no & no & yes & yes \\
irregular oval & no & no &  no & yes \\
\hline
\end{tabular}
\end{center}

\vspace*{0.2cm} \noindent Pursuing our analogy to the extreme, one
observes that whereas $d=2$ and $d=4$ can accommodate only one
kind of c-sets of MUBs, viz. those present also in odd dimensions
and having their counterparts in ordinary conics, $d=8$ should
already feature two different types and Hilbert spaces of $d\geq$
16 should be endowed with as many as three qualitatively different
kinds of such sets. So, if this analogy holds, a new MUBs' physics
is to be expected to emerge at the three-qubit level and become
fully manifested for four- and higher-order-qubit
states/configurations. 

Finally, we shall briefly address the non-Desarguesian case. We start with an observation 
that the definition of an oval is expressed in purely combinatorial terms and so it equally well
applies to finite {\it non}-Desarguesian  planes. These planes, however, do not admit
coordinatization in terms of any Galois field [14--16]; hence, the c-sets of MUBs corresponding to ovals
in such planes must fundamentally differ from ``Desarguesian" sets. The lowest order for which non-Desarguesian planes were found to exist is $d=9$, and there are even three distinct kinds of them; 
this means that it is also two-qutrit states whose properties merit a careful inspection.\footnote{It is a really intriguing fact to realize here that the two smallest non-trivial dimensions our approach singles out, viz. $d=8=2^3$ and $d=9=3^2$, are precisely those (product dimensions) where the so-called {\it unextendible product bases} (UPBs) first appear [see, e.g., 17,18]. This indicates that our oval geometries may underlie a wider spectrum of finite-dimensional quantum structures than sole MUBs.} 
The most tantalizing aspect of this analogy is, however, the case where
$d$ is composite (i.e., not a prime power) because such projective planes, if they exist, must
necessarily be non-Desarguesian [14,15]. So, if there exist c-sets of MUBs for $d$ composite, their
properties cannot be described in terms of {\it fields}; instead, one has to employ a more abstract 
concept, that of (planar) {\it ternary rings}, as these are proper systems for charting
non-Desarguesian projective planes [15,16]. And this is perhaps the most serious
implication of our approach and a serious challenge for further
geometrically-oriented explorations of MUBs, especially given an important role that MUBs start playing
in current quantum cryptographic schemes/protocols and quantum information theory in general.
\\\\
{\bf Acknowledgement}\\
The first author wishes to acknowledge the
support received from a 2004 ``S\'ejour Scientifique de Haut
Niveau" Physics Fellowship of the French Ministry of Youth,
National Education and Research (No. 411867G/P392152B).

\end{document}